# AI supported Topic Modeling using KNIME-Workflows[1]

Jamal Al Qundus[1], Silvio Peikert[1] and Adrian Paschke[1]

[1]Fraunhofer-Institut FOKUS, Berlin, Germany
{jamal.al.qundus, silvio.peikert, adrian.paschke}@fokus.fraunhofer.de

**Abstract.** Topic modeling algorithms traditionally model topics as list of weighted terms. These topic models can be used effectively to classify texts or to support text mining tasks such as text summarization or fact extraction. The general procedure relies on statistical analysis of term frequencies. The focus of this work is on the implementation of the knowledge-based topic modelling services in a KNIME[2] workflow. A brief description and evaluation of the DBPedia[3]-based enrichment approach and the comparative evaluation of enriched topic models will be outlined based on our previous work. DBpedia-Spotlight[4] is used to identify entities in the input text and information from DBpedia is used to extend these entities. We provide a workflow developed in KNIME implementing this approach and perform a result comparison of topic modeling supported by knowledge base information to traditional LDA. This topic modeling approach allows semantic interpretation both by algorithms and by humans.

## 1 Introduction

Recent developments related to Semantic Web made knowledge from the web available as machine readable ontologies. Links and vocabulary mappings between public ontologies enable algorithms to make use of knowledge from the web available as linked open data. One of the most popular public knowledge repositories is DBpedia. The DBpedia project extracts structured data from Wikipedia and makes it accessible as knowledge base via a SPARQL interface. ([1])

Topic modeling performs analysis on texts to identify topics. These topic models are used to classify documents and to support further algorithms to perform context adaptive feature, fact and relation extraction.

---

[1] This work has been partially supported by the "Wachstumskern Qurator – Corporate Smart Insights" project (03WKDA1F) funded by the German Federal Ministry of Education and Research (BMBF).

[2] https://www.knime.com/
[3] https://wiki.dbpedia.org/
[4] https://www.dbpedia-spotlight.org/

While Latent Dirichlet Allocation (LDA) [2], Pachinko Allocation [3], or Probabilistic Latent Semantic Analysis (PLSA) [4] traditionally perform topic modeling by statistical analysis of co-occurring words, the approaches in [1], [5] and [6] integrate semantics into LDA.

[1], [5] and [6] propose methods to improve word-based topic modeling approaches by introducing semantics from knowledge bases. This reduces perplexity issues arising from ambiguous terms and produces topic models that directly link to the knowledge base. Topic models created using a knowledge base are easier to understand by humans than topic models created exclusively by means of statistics.

This proof of concept work applies the method from [6] to perform knowledge base supported topic modeling using DBpedia. The presented approach to topic modeling is based on the semantics of entities identified in the document. The basic idea of LDA to perform analysis based on term frequency is maintained. The extension of [6] is to enrich the input using a knowledge base to perform LDA with semantics. Therefore, DBpedia Spotlight API is used to recognize entities and additional information to these entities is retrieved via the DBpedia API endpoint. During a preprocessing stage the text is tagged with semantic annotations from the knowledge base and the tagged text is used as input to the LDA algorithm. This results in improved topic models due to more context and less ambiguities in the input.

## 2    Architecture

The text to be examined is transferred to DBpedia Spotlight API. Spotlight returns a JSON object containing all entities recognized in the text. Additional information to these entities is retrieved using DBpedia API. The response for each entity is a set of properties e.g. *tags*, *Uri*, *type* and *hypernym* (see Section 4 for details). A tagger combines these sets with the corresponding entities in the text. The result is processed by LDA. LDA performs topic modeling and provides the result in two formats (as a table with weights and as an image visualization). The architecture of the processing pipeline is illustrated in **Fig. 1**.

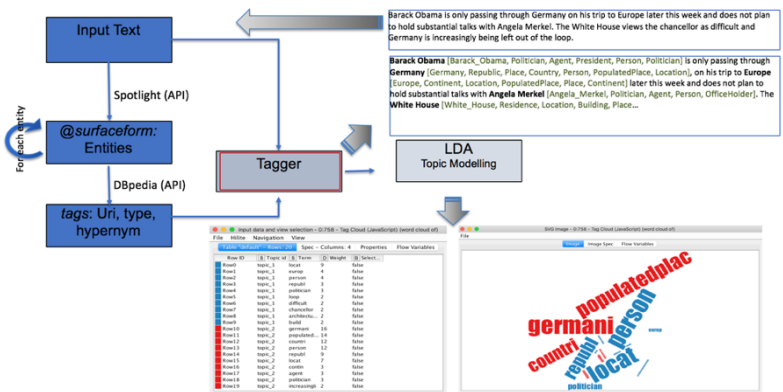

**Fig. 1** architecture of the topic modeling pipeline using DBpedia-Spotlight, DBpedia and the LDA algorithm.

## 3      KNIME

The KNIME information miner is an open source modular platform for visualization and selective execution of data pipelines. KNIME as a powerful data analysis tool that enables simple integration of algorithms, data manipulation and visualization methods in the form of modules or nodes [7].

There are a number of additional services in the KNIME ecosystem, e.g. KNIME Server, which connects the different actors (services, teams and individuals) in a central place and thus offers a platform for collaboration. KNIME Workflow Hub makes workflows publicly available on the KNIME Examples Server. Members of the user community can share workflows and receive ratings and comments from other users. In our work with the KNIME analytical platform we have implemented and performed various modelling methods to offer complete services around semantic analysis.

## 4      Workflow for Topic Modeling

The workflow developed in this work consists of four stages: (1) Reading the text in consideration and Entity recognition using DBpedia-Spotlight API. (2) Getting properties of the entities included in the JSON. (3) Tagging the text by combining the entities and the related properties gained from the previous phase. (4) Text cleaning and topic modeling using the LDA algorithm. **Fig. 2** gives an overview of the workflow developed and its modularization into four stages.

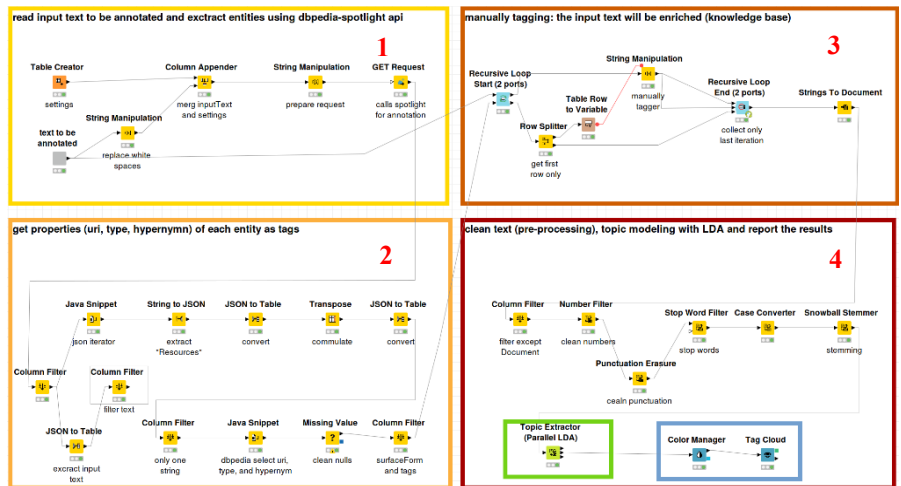

**Fig. 2** represents the workflow developed for topic modeling.

The Reading stage includes *Table Creator* node, which provides the settings of the parameters used to request entities from DBpedia Spotlight. We use confidence=0.5 and support=0 to get as many entities as possible from the text. *File Reader* node for reading the text from a path. *String Manipulation* node to repair the text, e.g. to replace double spaces. *Column Appender* node to combine the data provided by the *Table Creator* and *File Reader* nodes. *String Manipulation* node prepares the URL request to DBpedia Spotlight, which is then sent by the node *Get Request*. The output of this stage is a table of the text entities recognized by DBpedia Spotlight as shown in **Fig. 3**.

**Fig. 3** represents the response JSON object of DBpedia Spotlight.

The Get properties stage contains *Column Filter* node to extract the entities column from the table. *Java Snippet* node to filter *Resources* from the JSON. The *String to JSON*, *JSON to Table*, *Transpose* and *JSON to Table* nodes to put the column content in the format required for further processing. *Column Filter* node filters *types* and *surfaceForms*. *Java Snippet* node sends a HTTP request containing a SPARQL query to DBpedia API and retrieves entities and the related tags. *Missing Value* node deletes null values and *Column Filter* node filters *surface forms* (entities) and *tags* from the table created, which build the output of this stage as illustrated in **Fig. 4**.

**Fig. 4** represents the table filtered by surface name and tags

The tagging stage implements a loop taking the original input text and the recognized entities with their tags to match these entities with their mentions in the original text and enrich it by the tags as shown in **Fig. 5**. This loop consists of *Recursive Loop Start* to begin the loop, *Row Filter* to get the rows one by one, *Row Table to Variable* as a converter, *String Manipulation* as a tagger and *Recursive Loop End* to get back into the loop in case there are still entries in the table. At the end of the loop a *String to Document* node converts the text into a document format and forwards it to the next stage.

**Fig. 5** illustrates the tagged text

In the text cleaning and topic modeling stage, the produced document will be cleaned by *Column Filter*[5], *Number Filter*, *Punctuation Erasure*, *Stop Word Filter*, *Case Converter* and *Snowball Stemmer*. That preprocessed text is passed to *Topic Extractor* node that implements the LDA algorithm. LDA creates the topic model as list of weighted terms, which are then visualized using *Color Manager* and *Tag Cloud* nodes.

## 5      Evaluation

The focus of this work was on the implementation of the knowledge-based topic modelling services in a Knime workflow. For a detailed description and evaluation of the DBPedia based enrichment approach and the comparative assessment of enriched topic models we refer to our earlier work in [6] and [8]. In this paper we demonstrate the Knime-based proof-of-concept implementation by comparing the results of topic modeling supported by knowledge base information to traditional LDA using the following text:

*Barack Obama is only passing through Germany on his trip to Europe later this week and does not plan to hold substantial talks with Angela Merkel. The White House views the chancellor as difficult and Germany is increasingly being left out of the loop.*

This text is expanded with annotations from DBpedia as follows:

*Barack Obama [Barack_Obama, Politician, Agent, President, Person, Politician] is only passing through Germany [Germany, Republic, Place, Country, Person, PopulatedPlace, Location], on his trip to Europe [Europe, Continent, Location, PopulatedPlace, Place, Continent] later this week and does not plan to hold substantial talks with Angela Merkel [Angela_Merkel, Politician, Agent, Person, OfficeHolder]. The White House [White_House, Residence, Location, Building, Place…*

---

[5] *Column Filter* which is common needed, since the output of the most nodes includes, in addition to the result, its input that is mostly not needed any more.

**Fig. 6** shows the image visualization of weighted, normalized terms created by LDA without semantic annotations and **Fig. 7** the weighted, normalized terms obtained with support of a knowledge base.

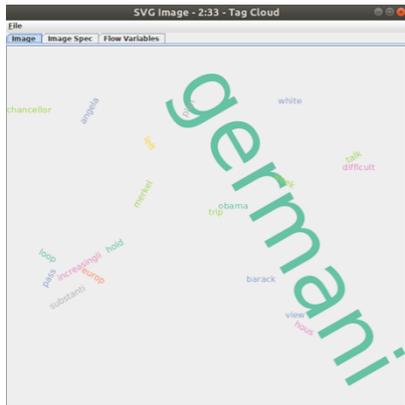
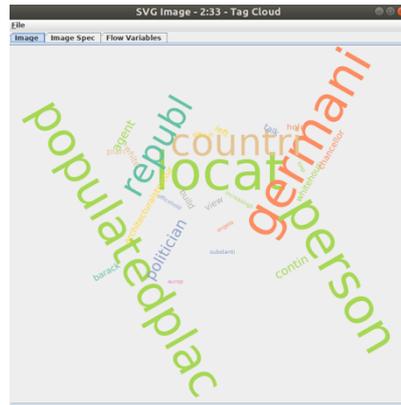

**Fig. 6** shows the traditional LDA topic model

**Fig. 7** shows the knowledge base enriched topic model

The results reflect the expectations. LDA provides a naive topic model for the original text comprising of weighted lemmatized terms from the input text with only one term having a significantly higher weight than other terms of the model. The knowledge base supported method creates a superior topic model also containing weighted lemmatized terms from the knowledge base, which are not present in the input text. This topic model enables semantic interpretation by algorithms as well as by humans.

In particular the enriched topic model enables algorithms to infer from the topic model linked to a knowledge base that the input text contains information about politics and actions of relevant officials, while a classification based on the traditional LDA topic model might result in a false classification as geographical text.

## 6  Summary and Future Prospects

This proof of concept work developed a KNIME workflow to perform comprehensive topic modeling using a knowledge base. The use of information from a knowledge base is achieved by using DBpedia Spotlight API for entity recognition and DBpedia API to retrieve entity properties. The presented results show that the developed approach is applicable and delivers results containing more comprehensive insights into a text than statistical topic models based on words only. The created topic models can improve the results of various methods used for text mining tasks such as text classification or fact and relation extraction.

Topic modeling using knowledge bases is a step towards improved automated methods for knowledge base population. Other methods in natural language processing might also be extendable by applying the idea of annotating text with information from knowledge bases. We expect improved results over word-based approaches for these tasks in future work, especially when analyzing small corpora.

**References**

[1] M. Allahyari and K. Kochut, 'Automatic topic labeling using ontology-based topic models', in *2015 IEEE 14th International Conference on Machine Learning and Applications (ICMLA)*, 2015, pp. 259–264.

[2] D. M. Blei, A. Y. Ng, and M. I. Jordan, 'Latent dirichlet allocation', *J. Mach. Learn. Res.*, vol. 3, no. Jan, pp. 993–1022, 2003.

[3] W. Li and A. McCallum, 'Pachinko allocation: DAG-structured mixture models of topic correlations', in *Proceedings of the 23rd international conference on Machine learning*, 2006, pp. 577–584.

[4] T. Hofmann, 'Probabilistic latent semantic analysis', in *Proceedings of the Fifteenth conference on Uncertainty in artificial intelligence*, 1999, pp. 289–296.

[5] I. Hulpus, C. Hayes, M. Karnstedt, and D. Greene, 'Unsupervised graph-based topic labelling using dbpedia', in *Proceedings of the sixth ACM international conference on Web search and data mining*, 2013, pp. 465–474.

[6] A. Todor, W. Lukasiewicz, T. Athan, and A. Paschke, 'Enriching topic models with DBpedia', in *OTM Confederated International Conferences" On the Move to Meaningful Internet Systems"*, 2016, pp. 735–751.

[7] M. R. Berthold *et al.*, 'KNIME-the Konstanz information miner: version 2.0 and beyond', *ACM SIGKDD Explor. Newsl.*, vol. 11, no. 1, pp. 26–31, 2009.

[8] Wojciech Lukasiewicz, Alexandru Todor, Adrian Paschke, '*Human Perception of Enriched Topic Models*'. In: Business Information Systems - 21st International Conference, BIS 2018, Berlin, Germany: 15-29